\documentstyle[psfig]{mn}
\begin{document}

\title{1999 Quadrantids and the lunar Na atmosphere}
\author[S. Verani et al.]
{S.~Verani$^1$\thanks{present address: Department of Astronomy, University of 
Padova}, C.~Barbieri$^2$, C.~R.~Benn$^3$, G.~Cremonese$^4$, M.~Mendillo$^5$\\
$^1$International Space Science Institute, Hallerstrasse 6, 3012 Bern,
Switzerland\\
$^2$Department of Astronomy, University of Padova, Vic. Osservatorio 2,
35122 Padova, Italy\\
$^3$Isaac Newton Group, Apartado 321, 38700 Santa Cruz de La Palma, Spain\\
$^4$Astronomical Observatory of Padova, Vic. Osservatorio 5, 35122 Padova,
Italy\\
$^5$Center for Space Physics, Boston University, Boston, Massachusetts 02215,
USA}
\maketitle

\begin{abstract}
Enhancements of the Na emission and temperature from the lunar atmosphere were
reported during the Leonids meteor showers of 1995, 1997 and 1998. Here we
report a search for similar enhancement during the 1999 Quadrantids, which have
the highest mass flux of any of the major streams. No enhancements were
detected. We suggest that different chemical-physical properties of the Leonid
and Quadrantid streams may be responsible for the difference.

\end{abstract}

\section{Introduction}
Evidence of a lunar atmopshere has been sought since the dawn of the telescopic
era, but none was found until {\it in situ} measurements by the Apollo missions
revealed the presence of He and Ar, both of solar origin, and possibly CH$_4$,
CO, CO$_2$, and N$_2$ (Hodges 1974). Discovery from the ground of Na
\cite{PoMo85} and K \cite{PoMo86} in Mercury's atmosphere renewed interest in
observing the lunar atmosphere from Earth, since both atmospheres are believed
to be generated by similar mechanisms, though at different rates. Their
atmospheres are continuosly lost and repopulated under the influence of the
interplanetary medium, and are known as {\it transient atmospheres}. There are
four candidate source mechanisms for such atmospheres: desorption and
sputtering by solar photons; chemical sputtering by the solar wind and by
energetic particles from the Earth's magnetosphere; micrometeoroid impacts; and
thermal desorption. The sink mechanisms are: Jeans escape; escape due to solar
radiation pressure; ionization by the solar wind; and photoionization. The
latter is the dominant sink mechanism for lunar Na, with proposed lifetimes of
15 or 47 hr at 1 AU (cf. Cremonese et al. 1997). The expected energy
distribution, characteristic temperature and spatial distribution of each
source mechanism is given by Smyth \& Marconi (1995).

Lunar Na and K were discovered in 1988 (Potter \& Morgan 1988; Tyler et al.
1988). Many observations using different techniques have been made to address
the origin and evolution of this atmosphere, at present thought a combination
of sputtering by solar photons and meteoroid impacts (Cremonese \& Verani 1997,
Sprague et al. 1998) or of chemical sputtering by solar wind ions and
photosputtering \cite{PoMo98} is proposed. For a comprehensive overview see
Stern (1999). Similar Na atmospheres are present around Jupiter's moons Io
\cite{Bro74} and Europa \cite{BroHi}, but the source mechanism for these is
believed to be sputtering by heavy ions in the Jovian magnetosphere. Na is a
good tracer of such thin atmospheres because of the ease with which its strong
emission can be observed.

\begin{table*}
\begin{center}
\begin{tabular}{ccrccc}
\noalign{\smallskip}\hline\hline
Date  & UT   &  \multicolumn{1}{c}{Side} & Scale Height  & Temperature    &  Brightness      \\ 
      &      &                           & km            & K              &  kR             \\  \hline
3 Jan & 3:24 & equator (w)               &  320 $\pm$ 49 & 1454 $\pm$ 223 & 0.83 $\pm$ 0.21 \\
      & 4:08 & north                     &  287 $\pm$ 41 & 1305 $\pm$ 187 & 0.86 $\pm$ 0.20 \\
      & 5:06 & north-west                &  272 $\pm$ 36 & 1238 $\pm$ 161 & 0.71 $\pm$ 0.17 \\ \hline
4 Jan & 0:36 & equator (w)               &  300 $\pm$ 40 & 1365 $\pm$ 185 & 0.76 $\pm$ 0.18 \\
      & 1:36 & north                     &  284 $\pm$ 38 & 1289 $\pm$ 171 & 0.70 $\pm$ 0.18 \\
      & 2:55 & north-west                &  305 $\pm$ 40 & 1388 $\pm$ 179 & 0.81 $\pm$ 0.19 \\
      & 3:46 & equator (e)               &       -       &        -       & 0.90 $\pm$ 0.21 \\
      & 4:34 & south                     &  345 $\pm$ 54 & 1572 $\pm$ 249 & 1.02 $\pm$ 0.24 \\ 
      & 5:36 & south-east                &  301 $\pm$ 42 & 1368 $\pm$ 193 & 0.79 $\pm$ 0.21 \\
\hline\hline\noalign{\smallskip}
\end{tabular}
\end{center}
\caption{Measurements of the lunar Na atmosphere during the Quadrantids. The
quoted errors are rms. They are larger than the intrinsic scatter between the
measurements due to systematic errors in the telescope's tracking of the Moon
during the 500 sec exposures.
The calculated temperature is defined as $mgH/k$, where $H$ is the real scale
height, $g$ is the Moon's acceleration of gravity (1.62 m s$^{-2}$), m is the
mass of the sodium atom (3.82 10$^{-23}$ g), and $k$ is Boltzmann's constant.
The tabulated brightnesses refer to the D2 line, 
and they are extrapolated
to the lunar surface. Observation of the east limb on Jan 4 yielded 
no measurements of
temperature and scale height.
The lunar phase angles on the first and second nights were 14$^o$
and 
26$^o$ respectively; 
the illuminated fractions of the Moon's disk were 0.98 and 0.95; and the
local solar zenith angles were 83$^o$ and 65$^o$. 
1 Rayleigh
(R)~=~10$^6$/4$\pi$~photons~cm$^{-2}$~s$^{-1}$~steradian$^{-1}$.}
\label{Osservazioni}
\end{table*}

Wide-angle imaging has shown that in the region between 1 and 10 lunar radii,
about 15\% of the Na atmosphere is due to micrometeor impacts \cite{Fl&Me95}.
Spectroscopic observations of the inner region of the atmosphere show that
meteor impacts have an effect which is spatially anisotropic, and which is
variable on short timescales (Cremonese \& Verani 1997; Sprague et al. 1998).
In 1991, Hunten, Kozlowski \& Sprague published the results of a 3-day campaign
of observations, during which the Na abundance increased by 60\% at
80$^{\circ}$ south, while that at the equator remained unchanged. They
suggested that the cause may have been a meteor shower, undetected by radar or
by other measurements, impacting near the south pole of the Moon. In 1998
Verani et al. reported a set of high-resolution spectra taken 4.5 days before
the maximum of the 1995 Leonids, with the same observational technique used for
the observations reported here, as described in Cremonese \& Verani (1997) and
in Sprague et al. (1992). They found significant enhancements of the brightness
and temperature (scale-height) of the Na atmosphere compared with previous
observations at similar lunar phase and local solar zenith angle. Since the
impact-generated component of the atmosphere has the highest expected
temperature \cite{Smy&Mar95}, the measured increase in temperature may be
associated with increased meteor activity during the Leonids.

A similar enhancement was detected during the maximum of the 1997 Leonids, at
Mount Lemmon and Asiago observatories \cite{Hun98}. These measurements appear
to be confirmed by observations taken with a different technique during the
night of the 1998 Leonids maximum \cite{Smi99}. The authors reported the
detection of a region of neutral Na emission in the direction of the
antisolar/lunar point in the 3 nights around new moon phase (18-20 November
1998), suggesting as the most likely cause the detection of the tail of the
Moon's atmosphere, i.e. its escaping component, driven outwards by solar
radiation pressure. A model of these results \cite{Wi99} indicates an increase
of a factor of 2 or 3 in the amount of Na escaping during the peak of the
Leonids.

Despite these observations, many issues remain to be resolved before one can
conclude with certainty that meteor showers are responsible for the observed
enhancements. In fact the lunar atmosphere shows time variations, the nature of
which is not fully understood (Sprague et al. 1998; Smith et al. 2000). A
measurement of enhanced Na emission from the Moon during a meteor shower other
than the Leonids would strongly support the hypothesis that these showers are
an important source of the transient Na atmosphere, both on the Moon and on
other solar-system bodies with thin atmospheres. In terms of mass flux
(g~cm$^{-2}$~s$^{-1}$), the Quadrantids are the most intense meteor stream; in
fact if we compare the mass flux of the meteor streams and the mass flux of the
sporadic meteoroids reported in Table~\ref{Parameters}, we can see the
micrometeoritic flux could increase up to 30\% - 50\% during the Quadrantids.
From this point of view, this stream gives a good opportunity to test the
impact mechanism; we therefore observed the lunar atmosphere during the 1999
Quadrantid meteor shower.

\section{Observations}

\begin{figure}
\centerline{\psfig{file=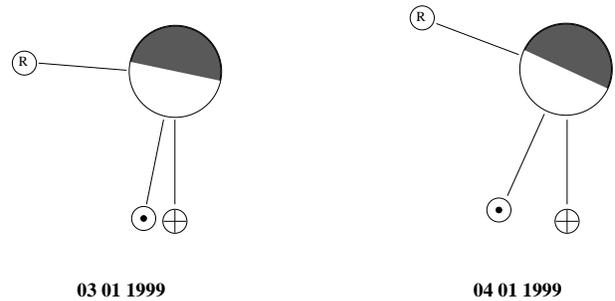,width=8truecm}}
\caption{The Moon as seen from the north ecliptic pole at UT~0$h$ on 3 and
4 January 1999. Shower maximum was during the latter night. The directions of
the Sun, the Earth and the radiant are shown. The radiant lay 70$^\circ$ above 
the plane of the ecliptic.}
\label{Configurazione}
\end{figure}

The observations were made in the early mornings of 3 and 4 January 1999 with
the Utrecht Echelle Spectrograph of the William Herschel Telescope on La Palma
(dispersion 0.053 \AA\ pixel$^{-1}$).  
Both nights were photometric. 
The Moon was full on Jan 2, at 02:49. The Sun was quiet. 
The diameter of the magnetotail is about 52 Earth radii at the Moon's
orbit \cite{ReRe75},
which means that the Moon was inside the magnetotail on the first night of the
observations (phase angle~=~14$^o$). On the second night (phase
angle~=~26$^o$) the Moon was near the predicted position of the magnetopause,
and thus its location with respect to magnetotail and magnetosheath is 
uncertain.

\begin{table*}
\begin{center}
\begin{tabular}{rcccccl}
\noalign{\smallskip}\hline\hline
\multicolumn{1}{c}{Date} 
                 & Ill. Frac. & LSZA & Distance from & Brightness                      & Temperature & Reference                    \\
                 &   \%       & deg  & surface (km)  &   kR                            &     K       &                              \\ \hline
21 Feb 1989 $^1$ & 0.99       & 90   &   50          &  1.25                           &     -       & Potter \& Morgan 1991        \\
22 Feb 1989 $^1$ & 0.98       & 83   &  10 - 70      &  1.18                           &     -       & Potter \& Morgan 1994        \\
02 Dec 1990 $^1$ & 0.99       & 87   &  10 - 70      &  0.90                           &     -       & Potter \& Morgan 1994        \\ 
04 Dec 1990 $^1$ & 0.96       & 71   &  10 - 70      &  1.17                           &     -       & Potter \& Morgan 1994        \\
22 Sep 1991 $^1$ & 0.96       & 71   &  10 - 70      &  0.17                           &     -       & Potter \& Morgan 1994        \\
23 Sep 1991 $^1$ & 0.99       & 87   &  10 - 70      &  0.27                           &     -       & Potter \& Morgan 1994        \\
24 Sep 1991 $^1$ & 0.99       & 87   &  10 - 70      &  0.78                           &     -       & Potter \& Morgan 1994        \\
20 Nov 1991 $^1$ & 0.95       & 65   &  10 - 70      &  0.43                           &     -       & Potter \& Morgan 1994        \\
29 Nov 1993 $^2$ & 1.00       & 90   &  2 - 12 R$_M$ &  0.90\rlap{$^\diamond$}         &     -       & Mendillo \& Baumgardner 1995 \\
06 Nov 1995 $^1$ & 1.00       & 90   &  50 - 180     &  1.38\rlap{$^\diamond$$^\star$} &   1538      & Cremonese \& Verani 1997     \\
03 Apr 1996 $^2$ & 1.00       & 90   &  2 - 12 R$_M$ &  0.82\rlap{$^\diamond$}         &     -       & Mendillo, Baumgardner \& Wilson 1999 \\
27 Sep 1996 $^2$ & 1.00       & 90   &  2 - 12 R$_M$ &  1.15\rlap{$^\diamond$}         &     -       & Mendillo, Baumgardner \& Wilson 1999 \\
24 Mar 1997 $^2$ & 1.00       & 90   &  2 - 12 R$_M$ &  1.80\rlap{$^\diamond$}         &     -       & Mendillo, Baumgardner \& Wilson 1999 \\  \hline
13 Oct 1990 $^1$ & 0.26       & 81   &  10 - 170     &  1.40\rlap{$^\diamond$}         &   1764      & Sprague et al. 1992                  \\
14 Oct 1990 $^1$ & 0.17       & 83   &  10 - 170     &  2.20\rlap{$^\diamond$}         &   1421      & Sprague et al. 1992                  \\
20 Nov 1991 $^1$ & 0.95       & 90   &  10 - 70      &  0.30                           &    -        & Potter \& Morgan 1994        \\
21 Nov 1991 $^1$ & 0.99       & 90   &  10 - 70      &  0.15                           &    -        & Potter \& Morgan 1994        \\
04 Dec 1990 $^1$ & 0.96       & 90   &  10 - 70      &  0.76                           &    -        & Potter \& Morgan 1994        \\
05 Dec 1990 $^1$ & 0.90       & 90   &  10 - 70      &  0.68                           &    -        & Potter \& Morgan 1994        \\
18 Sep 1995 $^1$ & 0.35       & 88   &   0 - 1800    &  6.13\rlap{$^\diamond$}         &   1332      & Sprague et al. 1998          \\
19 Sep 1995 $^1$ & 0.27       & 88   &   0 - 1800    &  1.06\rlap{$^\diamond$}         &   1451      & Sprague et al. 1998          \\ \hline
04 Jan 2000 $^1$ & 0.95       & 65   &  50 -  600    &  0.83\rlap{$^\diamond$}         &   1396      & This paper                   \\
\hline\hline\noalign{\smallskip}
\multicolumn{7}{l}{$^1$ high resolution spectroscopy} \\
\multicolumn{7}{l}{$^2$ wide angle imaging} \\
\multicolumn{7}{l}{$^\diamond$ value extrapolated at the surface} \\
\multicolumn{7}{l}{$^\star$ reanalysis of the data} \\
\multicolumn{7}{l}{R$_M$ (radius of the Moon) = 1736 km} \\
\end{tabular}
\end{center}
\caption{Previous measurements of Na emission from the lunar atmosphere. In the
upper part are listed the results of the observations carried out during near
full moon phase. In the bottom part are listed the results of the observations
taken near the polar zone (i.e. with local solar zenith angle similar to the
one of our observations) during any phase angle. A mean of the results reported
in this paper for the night of the Quadrantids maximum is shown in the last row
as comparison.}
\label{Confronto}
\end{table*}

The maximum of
the Quadrantids was predicted on the night Jan 3/4, with radiant at
$\alpha~=$~230$^\circ$, $\delta~=$~+49$^\circ$. The right ascensions of the
Moon at 0$^h$ UT on 3 and 4 January 1999 were respectively 115$^\circ$ and
129$^\circ$, so the shower fell on the Moon's western limb (cf.
Fig.~\ref{Configurazione}). The slit of the spectrograph, 150 arcsec long and
1 arcsec wide, was oriented perpendicular to the lunar limb, and observations
were made at various positions along the limb (Table~\ref{Osservazioni}) to
investigate changes in the Na emission with distance from the sub-radiant
point. Exposure time were typically 500 sec. The first 50 arcsec from the
surface had to be discarded due to the strong scattered moonlight; the
spatial coverage was then augmented placing two fields end-to-end. For details 
of the data reduction and analysis see Cremonese et al. (1992) and Verani et
al. (1998). 

The full moon prevented reliable visual measurements of the intensity of the
shower at the Earth, but 50-MHz radar observations made at Toyokawa Meteor
Observatory, Aichi, Japan, revealed an activity five times higher than sporadic
on Jan 3, between 20 and 24 UT, with a maximum of 235 events~hr$^{-1}$, in good
agreement with the observations of the 1997 and 1998 showers \cite{Suz}. In
addition, the MLT Dynamics Group of the University of Wales (V.~S.~Howells and
H.~R.~Middleton private communication) used the radar station at Aberystwyth,
UK, detecting a maximum on January 4, with activity two to three times higher
than the preceding and following nights. These observations suggest that the
1999 Quadrantids were at least as active as in previous years.

\section{Discussion}
The data were interpreted using Chamberlain's model of the exosphere
\cite{Chamberlain}, correcting the barometric density with an appropriately
adjusted partition function for the escaping and balistic components (cf.
Sprague et al. 1992). The results of the observations are shown in
Table~\ref{Osservazioni}. There are no significant differences between the
measurements made on Jan 3 and Jan 4, and none between measurements made in
directions towards the radiant and in other directions. The lack of
any observed change
between the first and second nights could also indicate
that there were no significant changes due to the Moon's exit from
the magnetotail; if indeed that took place.
While the differences between sputtering by magnetospheric and solar wind
particles are still not fully understood, our observations are consistent
with the dominant source mechanism being sputtering by solar radiation, which
will be isotropic around the limb at full moon 
(Mendillo \& Baumgartner 1995; Mendillo, Baumgartner \& Wilson 1999).

\begin{table*}
\begin{center}
\begin{tabular}{lccccc}
\noalign{\smallskip}\hline\hline
            & Impact velocity $^\ddagger$ & Total mass   & Mass flux                        & ZHR$_{max}$        & Active period (1998) \\
            &  km s$^{-1}$  & 10$^{12}$ kg & 10$^{-17}$ g cm$^{-2}$ s$^{-1}$  & \# h$^{-1}$        &   days               \\ \hline
Quadrantids & 41.0          & 1.3          &    3.3 \                         &  130               &  4                   \\
Geminids    & 34.4          & 16           &    2.4 \                         &  90                & 10                   \\
Perseids    & 59.4          & 31           &    0.28                          &  85                & 36                   \\
Orionids    & 66.4          & 3.3          &    0.18                          &  25                & 38                   \\
Leonids     & 70.7          & 6.7          &    0.3\rlap{$^{\star}$}          &  20\rlap{$^\star$} &  7                   \\
Sporadic    & 16.9          &              &    6.3\rlap{$^1$}                &                    &                      \\
            &               &              &    10.0\rlap{$^2$}               &                    &                      \\
            &               &              &    15.0\rlap{$^3$}               &                    &                      \\
\hline\hline\noalign{\smallskip}
\multicolumn{6}{l}{$^{\ddagger}$ average impact velocity on Earth}  \\
\multicolumn{6}{l}{$^{\star}$ the value increased up to 3 times
in 1995 and to 5 times in 1997 (Brown 1999)}  \\
\multicolumn{6}{l}{$^1$ sum over the mass range
10$^{-12}$ to 10$^{0}$~g (Gault, H\"orz \& Hartung 1972)} \\
\multicolumn{6}{l}{$^2$ sum over the mass range
10$^{-18}$ to 10$^{-2}$~g (Gr\"{u}n et al. 1985)} \\
\multicolumn{6}{l}{$^3$ sum over the mass range
10$^{-9}$ to 10$^{-4}$~g (Vanzani, Marzari \& Dotto 1997)} \\
\end{tabular}
\end{center}
\caption{Parameters at the Earth of five of the major meteor streams and of
sporadic meteors (Cook 1973; Hughes \&
McBride 1989; Love \& Brownlee 1993).}
\label{Parameters}
\end{table*}

In Table~\ref{Confronto} we compare measurements of the lunar Na atmosphere 
made by other authors, at full moon phase or at similar local solar zenith
angle with those reported here during the Quadrantids. The brightness and
temperature we measured during the Quadrantids are in good agreement with most
of the measurements made when there was no meteor shower, i.e. the 1999
Quadrantids had no detectable effect on the lunar Na atmosphere.

These results
may indicate that meteor impacts could have significant influence on the Moon's
atmosphere only under particular conditions, so that the detected difference
in the effects of the two streams may be due to physical differences between
the two streams. In fact the production of gas after an impact depends on
various factors, including the velocity, mass, and composition of the
impactors. The two streams differ in many parameters (cf.
Table~\ref{Parameters}), the impact velocity being one of the greatest: the
mean impact velocity of the Leonids (orbiting the Sun in the opposite direction
to the Earth and Moon) is 70.7~km~s$^{-1}$, much higher than that of the
Quadrantids (41.0~km~s$^{-1}$).

O'Keefe \& Ahrens (1977) have found that the amount of gas generated by an
impact is proportional to the factor S=($\rho_m$/$\rho_t$)(v/c$_p$)$^2$, where
v is the impact velocity, $\rho_m$ and $\rho_t$ are respectively the density of
the projectile and of the target, and c$_p$ is the bulk sound velocity in the
target. Starting from this model, Morgan, Zook \& Potter (1988) calculated the
amount of gas produced by impacts having impactor-target densities ratio of 1
and sound speed in the target of 7.44~km/s (estimated for Mercury's regolith
components). They found that a negligible amount of gas is produced at
velocities below 23 km/s (less than the mass of the projectile); at progressive
velocities this amount increases, becoming proportional to v$^2$, i.e. to the
kinetic energy, for velocities higher than 44 km/s, (cf. fig. 6 therein). In
our case (cometary grains into lunar regolith) the density ratio is $\leq$~1
(0.7 - 3 g/cm$^3$ vs 2.3 - 2.7 g/cm$^3$), so the threshold velocity should be
higher than the ones previously reported (up to 45 km/s).
Cintala (1992) in a model of high-velocity micrometeoroid impacts in the
regolith, also found that vapour production increases with impact velocity.
This model suggests that a difference of a factor 2 for the impact velocity
yields a factor of almost 3 in the amount of material melted, and a factor of
more than 4 in the amount of target material vaporised.

The effects of high-velocity impacts have also been investigated in the
laboratory (Eichhorn 1978), using a variety of materials for the targets
and for the impactors. The impact velocities ranged from 3 to 15~km/s, and
the masses from 10$^{-14}$ to 10$^{-9}$~g with a size of few microns; both
ranges are unfortunately lower than the typical ones for the Leonids and
Quadrantids. This experiment shows that the temperature of the generated gas
increases with the impact velocity (this could explain the enhancement of the
temperature of the Na atmosphere during the Leonids). These results seem to
confirm also the theoretical prediction of a low vapor production for such
impact velocities, as the mass of gas produced is less than twice the mass
of the impactor body. On the other hand, Schultz (1996) found that the
vaporization increases with the square (or even higher power)
of the velocity, 
even for velocities of 3-10~km/s, inconsistent with the hypothesis of 
a ``threshold'' velocity. The mean kinetic energies delivered
per unit area and time by the Leonids
(after 1994) and Quadrantids are of the same order of magnitude, so the
different impact velocities of the two streams may be responsible for the
observed differences only if the threshold velocity is real.

Different chemical compositions of the streams may affect the composition,
e.g., the Na abundance, of the gas produced. Spectroscopic measurements of the
airglow were made during the Leonid shower (Chu 2000, Nagasawa 1978),
suggesting a higher content of Na in Leonids with respect other meteoroids.
Moreover, Borovicka et al. (1999) found in their spectroscopic observations of
the Leonid meteors that smaller meteoroids tend to be poorer in Na than larger
ones, and a similar behaviour was found for the Perseids. \v{S}imek (1986)
investigated the mass distribution of five meteor streams (Geminids,
Quadrantids, Perseids, Leonids, and Giacobinids) and of the sporadics observed
with the radar. In his results he found an absence of larger particles in the
Quadrantids (and also in the sporadics). These results suggest that the Na
content of the Quadrantids is lower than that of the Leonids. Finally, age
could also play a role in depleting the content of volatiles in
Quadrantids meteoroids.  The Quadrantids are much
older than the Leonids, being created 7500 (Babadzhanov \& Ibrunov 1992) or 500
years ago (Jenniskens 1997).
In addition, because the Quadrantids have a shorter orbital period
(5.3 years,  vs 33 years for the Leonids), 
they have passed close to the Sun more
often, resulting 
in greater evaporation of volatile elements, such as Na, leaving the
meteoritic particles depleted in these elements.

\section{Conclusions}
No enhancement of the Na emission from the lunar atmosphere was detected
during the Quadrantids meteor shower. This contrasts with the reported
enhancements seen during at least three of the Leonids showers. As a possible
explaination we suggest the differences in the physical and chemical parameters
of the two streams, as widely explained in the previous section. To investigate
this hypothesis new measurements of hypervelocity impacts with a velocity range
resembling the one for the meteor showers and on a target more closely
approximating the the lunar regolith, or accurate modelling of such impacts,
are required. Measurements of the chemical composition of the meteoroids are
also required, as well as new observations of the lunar atmosphere during other
major meteor streams. For these reasons we believe that is not possible at the
moment to understand under which conditions the impact mechanism can generate
a significative amount of gas in the lunar atmosphere.

\section{Acknowledgements}
We are grateful to the referee, A.~Fitzsimmons, for valuable comments and
suggestions, and to J.~Geiss and to S.~H\aa land for useful discussions.
Barbieri acknowledges support from ASI;
Mendillo acknowledges support from the NASA Planetary Astronomy Program.
The WHT is operated on the island of La Palma by the Isaac Newton Group
in the Spanish Observatorio del Roque de los Muchachos of the Instituto
de Astrofisica de Canarias.

\end{document}